\documentclass[aps,pra,reprint, amsmath, amssymb,superscriptaddress,nofootinbib,crop=false]{revtex4-1}

\usepackage[subpreambles=true]{standalone}
\usepackage{import}
\usepackage{blindtext}

\usepackage{bm}
\usepackage[retainorgcmds]{IEEEtrantools}
\usepackage{graphicx}
\usepackage{mathrsfs}
\usepackage{amsmath}
\usepackage{amssymb}
\usepackage{color}
\usepackage{amsfonts}
\usepackage{times,txfonts}
\usepackage{nicefrac}
\usepackage[colorlinks=true,linkcolor=blue,urlcolor=blue,citecolor=blue,pdfusetitle]{hyperref}
\usepackage{amsmath}
\DeclareMathOperator{\sign}{sign}
\DeclareMathOperator{\sech}{sech}
\DeclareMathOperator{\arctanh}{atanh}

\newcommand{\tr}{\text{tr}}

\usepackage{subfiles} 

\begin{document}

\title{Symmetric Petz-Rényi relative entropy uncertainty relation}
\date{\today}
\author{Domingos S. P. Salazar}
\affiliation{Unidade de Educa\c c\~ao a Dist\^ancia e Tecnologia,
Universidade Federal Rural de Pernambuco,
52171-900 Recife, Pernambuco, Brazil}

\begin{abstract}
Holevo introduced a fidelity between quantum states that is symmetric and as effective as the trace norm in evaluating their similarity. This fidelity is bounded by a function of the trace norm, a relationship to which we will refer as Holevo’s inequality. More broadly, Holevo's fidelity is part of a one-parameter family of symmetric Petz-Rényi relative entropies, which in turn satisfy a Pinsker's-like inequality with respect to the trace norm. Although Holevo's inequality is tight, Pinsker's inequality is loose for this family. We show that the symmetric Petz-Rényi relative entropies satisfy a tight inequality with respect to the trace norm, improving Pinsker's and reproducing Holevo's as a specific case. Additionally, we show how this result emerges from a symmetric Petz-Rényi uncertainty relation, a result that encompasses several relations in quantum and stochastic thermodynamics.
\end{abstract}
\maketitle{}

{\bf \emph{Introduction -}} Holevo introduced a quantum fidelity \cite{Holevo1972} between states $\rho$ and $\sigma$ defined as
\begin{equation}
\label{fidelity}
F_H(\rho,\sigma) := [\tr (\sqrt{\rho}\sqrt{\sigma})]^2,
\end{equation}
 which is symmetric, $F_H(\rho,\sigma)=F_H(\sigma,\rho)$, and used under different names in the literature such as affinity and overlap \cite{Wilde2018,Iten2017,Zhihao2008,Albrecht1994,Shunlong2004,Sejong2013,Audenaert2014}. We have the following inequality for Holevo's fidelity,
\begin{equation}
\label{Holevos}
T(\rho,\sigma):=\frac{1}{2}|\rho - \sigma|_1 \leq \sqrt{1-F_H(\rho,\sigma)}, 
\end{equation}
where $|x|_1 :=\tr(\sqrt{x^\dagger x})$ is the trace norm. More generally, Holevo's fidelity is a member of a one parameter family of symmetric Petz-Rényi relative entropies defined as
\begin{equation}
\label{PetzRényi}
\tilde{D}_\alpha(\rho,\sigma):=\frac{1}{2}(D_\alpha(\rho||\sigma) + D_\alpha(\sigma||\rho)),
\end{equation}
such that $\tilde{D}_\alpha(\rho,\sigma)=\tilde{D}_\alpha(\sigma,\rho)$, where $D_\alpha(\rho||\sigma)$ is the Petz-Rényi relative entropy \cite{PETZ198657} defined as 
\begin{equation}
D_\alpha(\rho||\sigma):=\frac{1}{\alpha-1}\ln \tr(\rho^\alpha \sigma^{1-\alpha}),
\end{equation}
which is non negative for any $\alpha \in (0,1)\cup (1,\infty)$. It has applications in quantum hypothesis testing \cite{Datta2010,Brandao2011IEEE} and in quantum field theory \cite{Kudler2023}. The case $\alpha=1$ is understood as $D_1 (\rho,\sigma):=(D(\rho||\sigma)+D(\sigma||\rho))/2$, where $D(\rho||\sigma)=\tr(\rho\ln \rho)-\tr(\rho\ln\sigma)$ is the quantum relative entropy. For the specific case $\alpha=1/2$, one has $\tilde{D}_{1/2}(\rho,\sigma)=D_{1/2}(\rho||\sigma)=-2\ln \tr(\sqrt{\rho}\sqrt{\sigma})=-\ln F_H(\rho,\sigma)$ from (\ref{fidelity}) and (\ref{PetzRényi}). In this case, one could write Holevo's inequality (\ref{Holevos}) as a lower bound for $\tilde{D}_{1/2}(\rho,\sigma)$ as
\begin{equation}
\label{Holevos2}
\tilde{D}_{1/2}(\rho,\sigma) \geq \ln \frac{1}{1-T(\rho,\sigma)^2},
\end{equation}
also recently used in the applications of Petz-Rényi relative entropy in quantum field theory \cite{Kudler2023}, where it was explored a Pinsker's-like inequality,
\begin{equation}
\label{Pinskers}
\tilde{D}_\alpha (\rho,\sigma) \geq 2\min(\alpha,1)T(\rho,\sigma)^2.
\end{equation}
Note that, for the case $\alpha=1/2$, Holevo's inequality improves Pinsker's inequality for the particular case $\alpha=1/2$,
\begin{equation}
\label{Holevos3}
\tilde{D}_{1/2}(\rho,\sigma) \geq \ln \frac{1}{1-T(\rho,\sigma)^2} \geq T(\rho,\sigma)^2.
\end{equation}
Motivated by (\ref{Holevos3}), we analyze the following question: can we generalize Holevo's inequality (\ref{Holevos2}) beyond the case $\alpha=1/2$ such that it improves Pinsker's inequality for any $\alpha$ in the symmetric Petz-Rényi family? In this case, we obtain the following result:

({\it Generalized Holevo's inequality}) Let $\rho$ and $\sigma$ be any density matrices. Then,
\begin{equation}
\label{GenHolevos}
\tilde{D}_\alpha(\rho,\sigma) \geq \frac{1}{\alpha-1} \ln \frac{\cosh[(2\alpha-1)\arctanh(T(\rho,\sigma))]}{\cosh[\arctanh(T(\rho,\sigma))]}.
\end{equation}
For the specific case $\alpha=1/2$, our result (\ref{GenHolevos}) reproduces Holevo's inequality (\ref{Holevos2}). Notably, the case $\alpha=1$ previously appeared in the classic case ($[\rho,\sigma]=0$) as a bound for the symmetric Kullback-Leibler divergence in terms of total variation \cite{Dechant2022,Vo_2022,Salazar2022b}. We also show that (\ref{GenHolevos}) improves Pinsker's inequality (\ref{Pinskers}) for any $\alpha$. 

More generally, we show that (\ref{GenHolevos}) is a consequence of a quantum uncertainty relation for the symmetric Petz-Rényi relative entropy. This uncertainty relation not only results in (\ref{GenHolevos}), but it also results in multiple relations from stochastic and quantum thermodynamics as discussed in this letter. We state our main result:

{\bf \emph{Theorem:}} ({\it Symmetric Petz-Rényi uncertainty relation) Let $\rho$ and $\sigma$ be any density matrices and $\hat{\theta}$ be any Hermitian operator. Then,
\begin{equation}
\label{main}
\tilde{D}_\alpha(\rho,\sigma) \geq \frac{1}{\alpha-1} \ln \frac{\cosh[(2\alpha-1)\arctanh(s(\rho,\sigma;\hat{\theta}))]}{\cosh[\arctanh(s(\rho,\sigma;\hat{\theta}))]},
\end{equation}
where
\begin{equation}
s(\rho,\sigma;\hat{\theta}):=\Big[\frac{(1/2)(\langle \hat{\theta} \rangle_\rho - \langle \hat{\theta} \rangle_\sigma)^2}{\langle \langle \hat{\theta} \rangle \rangle_\rho + \langle \langle \hat{\theta} \rangle \rangle_\sigma + (1/2)(\langle \hat{\theta} \rangle_\rho - \langle \hat{\theta} \rangle_\sigma)^2}\Big]^{1/2},
\end{equation}}
where $\langle \hat{\theta}\rangle_{x}:=\tr\{x \hat{\theta}\}$ and $\langle \langle \hat{\theta}\rangle\rangle_x := \tr\{x \hat{\theta}^2\}-\tr\{x \hat{\theta}\}^2$. Note that the rhs in (\ref{main}) depends only on the statistics of $\hat{\theta}$ with respect to $\rho$ and $\sigma$ encoded in $s(\rho,\sigma;\hat{\theta})$, which is the main idea behind uncertainty relations discussed below. As a consequence of (\ref{main}), we obtain (\ref{GenHolevos}) and other results similar to thermodynamic uncertainty relations (TURs) \cite{Barato2015A,Gingrich2016,Polettini2017,Pietzonka2017,Hasegawa2019,Hasegawa2019b,VanTuan2020,Van_Vu_2020,Timpanaro2019B,Liu2020,Horowitz2022,Potts2019,Proesmans2019,Gianluca2022,Salazar2022d,Brandner2018,Carollo2019,Liu2019,VanVu2022b,Miller2021,Pires2021,Guarnieri2019,Hasegawa2019a,hasegawa2021,Hasegawa2023,Hasegawa2021b}. 

This letter is organized as follows. First, we present the main steps of the proof, then we show how the bound is saturated and the improvement over Pinsker's inequality. Finally, we discuss the applications of ($\ref{main}$) and obtain multiple previous and new results in stochastic and quantum thermodynamics.

{\bf \emph{Formalism:}}
The idea behind the proof is a strategy that maps $n$ dimensional quantum states $\rho = \sum_i p_i|p_i\rangle \langle p_i|$ and $\sigma=\sum_j q_j |q_j\rangle \langle q_j|$ into the following $n^2$ dimensional classic distributions: $P_{ij}:=p_i|\langle p_i|q_j \rangle |^2$ and $Q_{ij}:=q_j |\langle p_i | q_j \rangle|^2$. This strategy is sometimes called Nussbaum-Szkoła distributions \cite{Nussbaum2009} and it was successfully used in other contexts \cite{Datta2013,Salazar2023d}, particularly as tool to calculate relative entropies \cite{Androulakis2023,Androulakis2023b} as follows,
\begin{equation}
\label{formB1}
\sum_{ij}P_{ij}^\alpha Q_{ij}^{1-\alpha} = \sum_{ij}p_i^\alpha q_j^{1-\alpha}|\langle p_i|q_j \rangle|^2 = \tr(\rho^\alpha \sigma^{1-\alpha}),
\end{equation}
which results in $D_\alpha(P|Q) = D_\alpha(\rho||\sigma)$, where $D_\alpha(P|Q):=[1/(1-\alpha)]\ln \sum_{ij} P_{ij}^\alpha Q_{ij}^{1-\alpha}$ is the Rényi relative entropy, thus we have
\begin{equation}
\label{formB2}
\tilde{D}_\alpha(P,Q) = \tilde{D}_\alpha(\rho,\sigma),
\end{equation}
where $\tilde{D}_\alpha (P,Q)=(1/2)[D_\alpha (P|Q) + D_\alpha(Q|P)]$. Now that we mapped the Petz-Rényi relative entropy into a classic divergence, we import a recent result from information theory \cite{Nishiyama2022b},
\begin{equation}
\tilde{D}_\alpha (P,Q) \geq B(\alpha, \sqrt{\delta(P,Q)}),
\end{equation}
where $\delta(P,Q):=\sum_{s}(P_s-Q_s)^2/(P_s+Q_s)$ is the triangular discrimination and 
\begin{equation}
B(\alpha,x):=\frac{1}{\alpha-1} \ln \frac{\cosh[(2\alpha-1)\arctanh(x)]}{\cosh[\arctanh(x)]}
\end{equation}
is increasing in the interval $0< x < 1$ for all $\alpha>0$. Finally, we use the inequality $\sqrt{\delta(P,Q)} \geq s(\rho,\sigma, \hat{\theta})$ (see Appendix), which results in
\begin{equation}
\tilde{D}_\alpha(\rho,\sigma)=\tilde{D}_\alpha(P,Q) \geq B(\alpha,\sqrt{\delta(P,Q)})\geq B(\alpha,s(\rho,\sigma;\hat{\theta})),
\end{equation}
which proves our main result (\ref{main}). The generalized Holevo's inequality (\ref{GenHolevos}) is obtained as a particular case of (\ref{main}) where $\hat{\theta}=\sum_k \sign(w_k) |w_k \rangle \langle w_k|$ and $\{|w_k\rangle\}$ are the eigenvalues of $\rho-\sigma$ (see Appendix).

As in previous results, the bound (\ref{main}) is saturated for the specific two-level system, where $\rho=[e^{\epsilon/2}|1\rangle \langle 1| + e^{-\epsilon/2}|0\rangle \langle 0|]/(2\cosh(\epsilon/2))$,  $\sigma=[e^{-\epsilon/2}|1\rangle \langle 1| + e^{\epsilon/2}|0\rangle \langle 0|]/(2\cosh(\epsilon/2))$ and $\hat{\theta}=\phi(|1\rangle \langle 1| - |0\rangle \langle 0|)$. In this case, one has $\tr(\rho \hat{\theta})=\phi \tanh(\epsilon/2)$,  $\tr(\sigma\hat{\theta})=-\phi \tanh(\epsilon/2)$ and $\tr(\rho \hat{\theta}^2)=\tr(\sigma \hat{\theta}^2)=\phi^2$, such that
\begin{equation}
\tilde{D}_\alpha (\rho,\sigma) = \frac{1}{1-\alpha} \ln \frac{\cosh[(2\alpha-1)\epsilon/2]}{\cosh[\epsilon/2]},
\end{equation}
and also $s(\rho,\sigma;\hat{\theta})=|\tanh(\epsilon/2)|$. Therefore, in this minimal system, we have the saturation of (\ref{main}),
\begin{equation}
\tilde{D}_\alpha(\rho,\sigma) = B(\alpha,s(\rho,\sigma;\hat{\theta})).
\end{equation}
We also show that (\ref{GenHolevos}) improves Pinsker's inequality (\ref{Pinskers}) in the Appendix, where we have
\begin{equation} 
\tilde{D}_\alpha (\rho,\sigma) \geq B(\alpha,T(\rho,\sigma)) \geq 2\min(\alpha,1)T(\rho,\sigma)^2,
\end{equation}
which generalizes Holevo's case (\ref{Holevos3}) for any $\alpha$.

{\bf \emph{Discussion - }} We investigate particular cases of (\ref{main}) and (\ref{GenHolevos}) in stochastic and quantum thermodynamics. Some of the cases are very well known and some of them are new to our knowledge. 

First, turning our attention to relation (\ref{GenHolevos}), we note that particular cases also appeared in the literature. The case $\alpha=1/2$ is obviously the {\it Holevo's inequality} (\ref{Holevos2}), but we also observe that the case $\lim \alpha\rightarrow 1$ results in
\begin{equation}
\label{disc1}
\tilde{D}(\rho,\sigma) \geq 2T(\rho,\sigma)\arctanh(T(\rho,\sigma)),
\end{equation}
which was recently used to to analyse fluxes in quantum thermodynamics \cite{Salazar2024a}. We also note that the classic case ($[\rho,\sigma]=0$) of (\ref{disc1}) was used in the study of Markov chains in stochastic thermodynamics as well \cite{Dechant2022,Vo_2022,Salazar2022b}, where $\rho$ and $\sigma$ can be written as classic probabilities $p=(p_1,...,p_n)$, $q=(q_1,...,q_n)$, yielding
\begin{equation}
\label{TVbound}
\tilde{D}(p,q)\geq 2 \Delta(p,q) \arctanh(\Delta(p,q)),
\end{equation}
where $\Delta(p,q)=(1/2)\sum_s |p_i-q_i|$ is the total variation.

Moreover, we note that (\ref{main}) can be inverted to the following expression when $\langle \hat{\theta}\rangle_\rho \neq \langle \hat{\theta}\rangle_\sigma$,
\begin{equation}
\label{alphaqTUR}
\frac{\langle \langle \hat{\theta} \rangle \rangle_\rho + \langle \langle \hat{\theta} \rangle\rangle_\sigma}{(1/2)(\langle \hat{\theta} \rangle_\rho- \langle \hat{\theta} \rangle_\sigma)^2} \geq f(\alpha, \tilde{D}_\alpha(\rho,\sigma)),
\end{equation}
where $f(\alpha,x):=1/[B^{-1}(\alpha,D)]^2 -1$, and for a fixed $\alpha$, $B^{-1}(\alpha,x)$ is the inverse of $B(\alpha,x)$ for $x\geq0$, such that $B^{-1}(\alpha,B(\alpha,x))=x$. Remarkably, expression (\ref{alphaqTUR}) has the form of a quantum uncertainty relation, which explains the name of the theorem as {\it symmetric Petz-Rényi uncertainty relation}. As matter of fact, the case $\alpha=1$ results in the recently proposed {\it quantum relative entropy uncertainty relation} \cite{Salazar2023d},
\begin{equation}
\label{qTUR}
\frac{\langle \langle \hat{\theta} \rangle \rangle_\rho + \langle \langle \hat{\theta} \rangle\rangle_\sigma}{(1/2)(\langle \hat{\theta} \rangle_\rho- \langle \hat{\theta} \rangle_\sigma)^2} \geq f(1, \tilde{D}(\rho,\sigma)),
\end{equation}
where $f(1,x)=1/\sinh^2(g(x)/2)$ and $g(x)$ is the inverse of $h(x)=x\tanh(x/2)$ for $x\geq0$.  Analogously, the case $\alpha=1/2$ in (\ref{alphaqTUR}) results in the following {\it Holevo's uncertainty relation},
\begin{equation}
\label{HolevosqUR}
\frac{\langle \langle \hat{\theta} \rangle \rangle_\rho + \langle \langle \hat{\theta} \rangle\rangle_\sigma}{(1/2)(\langle \hat{\theta} \rangle_\rho- \langle \hat{\theta} \rangle_\sigma)^2} \geq \frac{F_H(\rho,\sigma)}{1-F_H (\rho,\sigma)},
\end{equation}
which takes the usual form of thermodynamic uncertainty relations as a lower bound for some uncertainty in terms of a dissimilarity. In this case, the dissimilarity is not the usual entropy production, but given in terms of the Holevo's fidelity (\ref{fidelity}) instead.

We also consider the classic situation $[\rho,\sigma]=0$, which represents the absence of coherence between the states. In this particular case, writing again $\rho$ and $\sigma$ in terms of classic probabilities $p=(p_1,...,p_n)$, $q=(q_1,...,q_n)$ and $\theta=(\theta_1,...,\theta_n)$ is a random variable. We obtain the classic version of (\ref{alphaqTUR}),
\begin{equation}
\label{alphaiTUR}
\frac{\langle \langle \theta \rangle \rangle_p + \langle \langle \theta \rangle\rangle_q}{(1/2)(\langle \theta \rangle_p - \langle \theta \rangle_q)^2} \geq f(\alpha, \tilde{D}_\alpha(p,q)),
\end{equation}
which can be seen as the $\alpha-$generalized version of the tightest form of the {\it hysteretic thermodynamic uncertainty relation} ($\alpha=1$) \cite{Proesmans2019,Gianluca2022,Salazar2022d},
\begin{equation}
\label{iTUR}
\frac{\langle \langle \theta \rangle \rangle_p + \langle \langle \theta \rangle\rangle_q}{(1/2)(\langle \theta \rangle_p - \langle \theta \rangle_q)^2} \geq f(1, \tilde{D}(p,q)),
\end{equation}
while the case $\alpha=1/2$ in (\ref{alphaiTUR}) 
can be written as (\ref{HolevosqUR}) but in terms of the Bhattacharyya or Hellinger distances.

Finally, in the particular case $p=P(\Gamma)$, $q=P(\Gamma^\dagger)$, where $\Gamma$ typically represents a trajectory and $\Gamma^\dagger$ is the inverse trajectory, such that $(\Gamma^\dagger)^\dagger = \Gamma$, and $\theta(\Gamma)$ is a current with property $\theta(\Gamma^\dagger)=-\theta(\Gamma)$, relation (\ref{iTUR}) yields a result known as the {\it thermodynamic uncertainty relation from the exchange fluctuation theorem} \cite{Hasegawa2019,Timpanaro2019B},
\begin{equation}
\label{xTUR}
\frac{\langle \langle \theta \rangle \rangle_p}{\langle \theta \rangle_p^2} \geq f(1, \langle \Sigma \rangle),
\end{equation}
where $\langle \Sigma \rangle := \sum_\Gamma P(\Gamma) \ln P(\Gamma)/P(\Gamma^\dagger)$ is the average entropy production.

{\bf \emph{Conclusions - }}
We proposed a symmetric Petz-Rényi uncertainty relation (\ref{main}) and studied multiple applications. Our result was obtained using a mapping from quantum to classic systems and exploring recent results from information theory. From our result, we wrote a general uncertainty relation (\ref{alphaqTUR}) and, for the particular case $\alpha=1/2$, we obtained a uncertainty relation in terms of Holevo's fidelity (\ref{qTUR}), a generalized Holevo's inequality (\ref{GenHolevos}) and multiple known results in quantum and stochastic thermodynamics (\ref{TVbound},\ref{iTUR},\ref{xTUR}). Our results highlight that different symmetric quantum dissimilarities may also play a role akin to the entropy production in expressions resembling thermodynamic uncertainty relations. This fact does not depend on specific properties of the system (such as the detailed fluctuation theorem), but it is rather a fundamental interplay between quantum uncertainties and symmetric dissimilarities.

{\bf \emph{Appendix- }}
We start proving a result in information theory for probabilities ($P,Q$). Then, we will show the quantum case is obtained as a consequence of the classic case using a strategy that maps $n$dimensional quantum states into $n^2$ dimensional classic distributions.

{\bf \emph{Definition 1.}}
Let $P=\{P_s\}$ and $Q=\{Q_s\}$ be probabilities in a set $S$, and let be the triangular discrimination be defined as
\begin{equation}
\label{def11}
\delta(P,Q):=\sum_s \frac{(P_s-Q_s)^2}{(P_s + Q_s)},
\end{equation}
with the notation $0^2/0=0$ for the cases where $P_s=Q_s$. We define the Rényi relative entropy as 
\begin{equation}
\label{def12}
D_\alpha(P|Q):=\frac{1}{\alpha-1}\ln \sum_s P_s^\alpha Q_s^{1-\alpha},
\end{equation}
for $\alpha \in (0,1)\cup(1,\infty)$ and the symmetric version,
\begin{equation}
\label{def13}
\tilde{D}_\alpha(P,Q):=\frac{1}{2}(D_\alpha(P|Q)+D_\alpha (Q|P)).
\end{equation}

{\bf \emph{Lemma 1.}} Let $P$ and $Q$ be any distributions, then
\begin{equation}
\label{Lemma1}
\tilde{D}_\alpha (P,Q) \geq  \frac{1}{\alpha-1} \ln \frac{\cosh[(2\alpha-1)\arctanh(\sqrt{\delta(P,Q)})]}{\cosh[\arctanh(\sqrt{\delta(P,Q)}]}.
\end{equation}

{\it Proof.} This is a particular case of a recent result for $f-$divergences \cite{Nishiyama2022b}. We provide an alternative poof using a formulation in terms of stochastic entropy \cite{Salazar2022d,Salazar2023a}. Let $\{P_s\}$ be a probability function and let $s':=m(s)$ be any involution (such that $m(m(s))=s$) and define a new probability $P'(s):=P(s')$. Unless mentioned otherwise, expectations $\langle \rangle$ are meant with respect to $P_s$, $\langle \rangle = \langle \rangle_P$. If the pair ($P,P'$) is not absolutely continuous (ie, there is an $s$ such that $P_s'=0$ and $P_s>0$), then $\tilde{D}_\alpha (P,Q) = \infty$ and (\ref{Lemma1}) holds immediately. Thus, we focus on the cases where $(P,P')$ are absolutely continuous ($P_s = 0 \leftrightarrow P_s'=0$, for all $s\in S$). Let $\Sigma(s)$ be an entropy-like random variable defined as 
\begin{equation}
\label{l1form1}
\Sigma(s):=\ln \frac{P(s)}{P'(s)},
\end{equation}
when $P_s'\neq 0$ and $\Sigma(s)=0$ otherwise. In the stochastic thermodynamics literature, this is known as the strong detailed fluctuation theorem and $\Sigma$ is the stochastic entropy production, although in our notation (\ref{l1form1}) is just a definition of a random variable $\Sigma(s)$. Consider the following expectation with respect to $P$,
\begin{equation}
\label{l1form2}
\langle \exp(\beta \Sigma) \rangle = \sum_s P(s)^{\beta+1} P'(s)^{-\beta},
\end{equation}
which can also be rewritten as $\exp(\beta\Sigma)=\sinh(\beta\Sigma) + \cosh(\beta\Sigma)$. Now using the following property, $\langle u(\Sigma)\rangle = \langle u(\Sigma)\tanh(\Sigma/2)\rangle$ for odd functions $u(-x)=-u(x)$, we obtain
\begin{equation}
\label{l1form3}
\langle \exp(\beta\Sigma) \rangle = \langle \sinh(\beta \Sigma)\tanh(\Sigma/2) + \cosh(\beta\Sigma)\rangle,
\end{equation}
which has the following compact form from $\cosh(x+y)=\cosh(x)\cosh(y)+\sinh(x)\sinh(y)$ and replacing $\beta=\alpha-1$,
\begin{equation}
\label{l1form4}
\langle \exp((\alpha-1)\Sigma) \rangle = \langle \frac{\cosh[(\alpha-1/2)\Sigma]}{\cosh(\Sigma/2)}\rangle.
\end{equation}
Now we note that the triangular discrimination $\delta(P,P')$ has a particular form in terms of the statistics of $\Sigma(s)$ as
\begin{equation}
\label{l1form41}
\delta(P,P')=\sum_s \frac{(P_s-P_s')^2}{(P_s+P_s')^2}\frac{P_s+P_s'}{2} = \langle \tanh(\Sigma/2)^2 \rangle,
\end{equation}
using the fact that $\langle \tanh(\Sigma/2)^2 \rangle_{(P+P')/2}=\langle \tanh(\Sigma/2)^2\rangle_P$.
Finally, we write the function $F(\Sigma):=\cosh[(\alpha-1/2)\Sigma]/\cosh(\Sigma/2)$ from (\ref{l1form41})  in terms of $\tanh(x/2)^2$
\begin{equation}
\label{l1form5}
F(\Sigma)=F(2\arctan(\sqrt{\tanh(\Sigma/2)^2})),
\end{equation}
and use Jensen's inequality in (\ref{l1form5})
\begin{eqnarray}
\label{l1form6}
   \langle F(\Sigma) \rangle \geq F(2\arctan(\sqrt{\langle \tanh(\Sigma/2)^2 \rangle}))=
    \nonumber
    \\
   = F(2\arctanh(\sqrt{\delta(P,P')}),
\end{eqnarray}
since $d^2 F(2\arctanh(\sqrt{y}))/dy^2 \geq 0$ for $\alpha>1$. Similarly, we get $\langle F(\Sigma) \rangle \leq F(2\arctan(\sqrt{\langle \tanh(\Sigma/2)^2 \rangle}))$ for $0\leq \alpha < 1$. Combining (\ref{l1form4}) and (\ref{l1form6}), we obtain
\begin{equation}
\label{l1form7}
\langle \exp(\alpha-1) \Sigma \rangle \geq \frac{\cosh[(2\alpha-1)\arctanh(\sqrt{\delta(P,P')}]}{\cosh(\arctanh(\sqrt{\delta(P,P')})},
\end{equation}
for $\alpha > 1$ and $\langle \exp(\alpha-1) \Sigma \rangle \leq F(2\arctanh (\sqrt{\delta(P,Q)}))$ for $\alpha < 1$. Note that Rényi relative entropy is given by
\begin{equation}
\label{l1form8}
D_\alpha(P|P') = \frac{1}{\alpha-1} \ln \langle \exp(\alpha-1)\Sigma \rangle,
\end{equation}
so that the cases $\alpha>1$ in (\ref{l1form8}) and the case $\alpha<1$ can be rewritten as a single expression,
\begin{equation}
\label{l1form9}
D_\alpha(P|P') \geq \frac{1}{\alpha-1} \ln \frac{\cosh[(2\alpha-1)\arctanh(\sqrt{\delta(P,P')})]}{\cosh(\arctanh(\sqrt{\delta(P,P')})},
\end{equation}
for all $\alpha \neq 1$. The case $\alpha=1$ should be understood as the limit $\alpha \rightarrow 1$. Since (\ref{l1form9}) was proved for any probability $P$ and any involution $m(s)$, we apply the expression for the particular set $\{(s,i)\}$, where $s \in S$ and $i \in \{0,1\}$, with involution $m(s,i)=(s,m(i))$, where $m(i)=1-i$ and probabilities $p(s,1):=P(s)/2$ and $p(s,0)=Q(s)/2$.  In this case,
we obtain $D_\alpha (p|p') = (1/2)(D_\alpha(P|Q)+D_\alpha(Q|P))= \tilde{D}_\alpha (P,Q)$ and $\delta(p,p')=\delta(P,Q)$, which results in our Lemma 1 (\ref{Lemma1}) for any $P,Q$. $\blacksquare$

{\bf \emph{Lemma 2.}} Let $\Theta_s \in \mathbb{C}$ be a complex random variable, then
\begin{equation}
\label{Lemma2}
\delta(P,Q) \geq \frac{(1/2)|\langle \Theta \rangle_P - \langle \Theta \rangle_Q|^2}{\langle \langle \Theta \rangle \rangle_P + \langle \langle \Theta \rangle \rangle_Q + (1/2)|\langle \Theta \rangle_P - \langle \Theta \rangle_Q|^2}, 
\end{equation}
where $\langle \langle \Theta \rangle \rangle := \langle |\Theta|^2 \rangle - |\langle \Theta \rangle|^2$.

{\it Proof.} This a is a complex generalization of a previous result \cite{Falasco_2022} also used in \cite{Salazar2024a} and we reproduce here. 
Consider probabilities $P,Q$ in $s \in S$, $\sum_s P(s)= \sum_s Q(s)=1$ and a complex valued random variable $\Theta(s) \in \mathbb{C}$. We define $S'=\{s \in S| P(s)+Q(s)>0\}$ and the probability $\tilde{P}(s):=(P(s)+Q(s))/2$ in $S'$, $\sum_{s \in S'}\tilde{P}(s)=1$, $\overline{\Theta}_X:=\langle \Theta \rangle_X=\sum_s \Theta(s) X(s)$, for $X \in \{P,Q,\tilde{P}\}$. 
Note that the expression $|\overline{\Theta}_P - \overline{\Theta}_Q|^2$ can be rewritten as
\begin{equation}
\label{app1}
\frac{1}{4}|\overline{\Theta}_P - \overline{\Theta}_Q|^2 = |\sum_{s\in S'} (\Theta(s)-c) \frac{(P(s)-Q(s))}{2}|^2,
\end{equation}
for any complex $c$. Using Cauchy-Schwarz inequality, we also obtain for any complex $c$,
\begin{equation}
\label{app2}
|\sum_{s\in S'} (\Theta(s)-c) \frac{(P(s)-Q(s))}{2}|^2 \leq \langle |\Theta-c|^2 \rangle_{\tilde{P}}\langle (\frac{P-Q}{P+Q})^2\rangle_{\tilde{P}},
\end{equation}
so that combining (\ref{app1}) and (\ref{app2}) for $c=\overline{\Theta}_{\tilde{P}}$, it yields 
\begin{equation}
\label{app3}
\frac{1}{4}|\overline{\Theta}_P - \overline{\Theta}_Q|^2 \leq \langle |\Theta-\overline{\Theta}_{\tilde{P}}|^2 \rangle_{\tilde{P}}\langle (\frac{P-Q}{P+Q})^2\rangle_{\tilde{P}}.
\end{equation}
We note from Definition 1 (\ref{def11}) that
\begin{equation}
\label{app4}
\delta(P,Q)=\langle (\frac{P-Q}{P+Q})^2\rangle_{\tilde{P}}.
\end{equation}
Now consider the identity
\begin{equation}
\label{app5}
4\langle |\Theta-\overline{\Theta}_{\tilde{P}}|^2 \rangle_{\tilde{P}} = 2(\langle|\Theta|^2\rangle_P-|\overline{\Theta}_P|^2) + 2(\langle|\Theta|^2\rangle_Q-|\overline{\Theta}_Q|^2) + |\overline{\Theta}_P - \overline{\Theta}_Q|^2. 
\end{equation}
Combining (\ref{app3}), (\ref{app4}) and (\ref{app5}) it results in (\ref{Lemma2}) $\blacksquare$.

{\bf \emph{Definition 2.}} ({\it Nussbaum-Szkoła distributions}) Let $\rho$ and $\sigma$ be density matrices in $n$ dimensions with spectral decomposition $\rho = \sum_i^n p_i |p_i\rangle \langle p_i|$ and $\sigma=\sum_j^n q_j |q_j \rangle \langle q_j|$. We define distributions $P$ and $Q$ in $S=\{(i,j)|1\leq i\leq n; 1\leq j \leq n\}$ as 
\begin{equation}
P_{ij}:=|\langle p_i | q_j \rangle|^2 p_i,
\end{equation}
\begin{equation}
Q_{ij}:=|\langle p_i | q_j \rangle|^2 q_j,
\end{equation}
and we add the definition of the auxiliary complex random variable $\Theta$ based on a Hermitian operator $\hat{\theta}$,
\begin{equation}
\Theta_{ij} := \frac{\langle p_i | \hat{\theta} | q_j \rangle} {\langle p_i | q_j \rangle},
\end{equation}
for $\langle p_i | q_j \rangle \neq 0$ and $\Theta_{ij}:=0$ otherwise.

{\bf \emph{Lemma 3.}} For any density matrices $\rho, \sigma$ and Hermitian operator $\hat{\theta}$ we have
\begin{eqnarray}
\label{Lemma3}
\frac{(1/2)|\langle \Theta \rangle_P - \langle \Theta \rangle_Q|^2}{\langle \langle \Theta \rangle \rangle_P + \langle \langle \Theta \rangle \rangle_Q + (1/2)|\langle \Theta \rangle_P - \langle \Theta \rangle_Q|^2} \nonumber
\\ \geq
\frac{(1/2)(\langle \hat{\theta} \rangle_\rho - \langle \hat{\theta} \rangle_\sigma)^2}{\langle \langle \hat{\theta} \rangle \rangle_\rho + \langle \langle \hat{\theta} \rangle \rangle_\sigma + (1/2)(\langle \hat{\theta} \rangle_\rho - \langle \hat{\theta} \rangle_\sigma)^2},
\end{eqnarray}
with $P,Q,\Theta$ given by Definition 2 and $\langle \hat{\theta}\rangle_{\rho}:=\tr\{\rho \hat{\theta}\}$, $\langle \hat{\theta}\rangle_\sigma := \tr\{\sigma \hat{\theta}\}$, $\langle \langle \hat{\theta}\rangle\rangle_\rho := \tr\{\rho \hat{\theta}^2\}-\tr\{\rho \hat{\theta}\}^2$ and $\langle \langle \hat{\theta}\rangle\rangle_\sigma := \tr\{\sigma \hat{\theta}^2\}-\tr\{\sigma \hat{\theta}\}^2$.

{\it Proof.} This idea was explored in \cite{Salazar2023d}. The expected value of $\hat{\theta}$ with respect to $\rho$ is
\begin{equation}
\label{l3form1}
\tr(\rho \hat{\theta})  = \sum_{ij} p_i \langle p_i | \hat{\theta}|q_j\rangle \langle q_j | p_i \rangle = \sum_{ij; \langle q_j |p_i \rangle \neq 0} p_i |\langle q_j | p_i \rangle|^2 \frac{\langle p_i | \hat{\theta}|q_j\rangle}{\langle p_i |q_j \rangle},
\end{equation}
where we used $\langle p_i | q_j \rangle = \langle q_j | p_i \rangle^*$. Using Definition 2, in terms of $P$ and $\Theta$, we have from (\ref{l3form1}),
\begin{equation}
\label{l3form3}
\tr(\rho \hat{\theta}) = \sum_{ij}P_{ij}\Theta_{ij} := \langle \Theta \rangle_P.
\end{equation}
Similarly, we obtain for the expected value of $\hat{\theta}$ with respect to $\sigma$ using Definition 2,
\begin{equation}
\label{l3form4}
\tr(\sigma \hat{\theta}) = \sum_{ij}Q_{ij}\Theta_{ij} := \langle \Theta \rangle_Q.
\end{equation}
Analogously, we have for the expected value of $\hat{\theta}^2$ with respect to $\rho$,
\begin{eqnarray}
\label{l3form5}
\tr(\rho \hat{\theta}^2) = \sum_{ij} p_i |\langle p_i |\hat{\theta}|q_j \rangle|^2  \geq 
\\
\sum_{ij;\langle q_j|p_i\rangle \neq 0} p_i |\langle p_i |\hat{\theta}|q_j \rangle|^2 = \sum_{ij} P_{ij}|\Theta_{ij}|^2,
\end{eqnarray}
where we used $\hat{\theta}=\hat{\theta}^\dagger$, which yields 
\begin{equation}
\label{l3form7}
\tr(\rho \hat{\theta}^2) \geq \sum_{ij} P_{ij}|\Theta_{ij}|^2 :=\langle |\Theta|^2 \rangle_P.
\end{equation}
We have a similar expression in terms of $\sigma$,
\begin{equation}
\label{l3form8}
\tr(\sigma \hat{\theta}^2) \geq \sum_{ij} Q_{ij}|\Theta_{ij}|^2 :=\langle |\Theta|^2 \rangle_Q.
\end{equation}
Combining expressions (\ref{l3form3}), (\ref{l3form4}), (\ref{l3form7}) and (\ref{l3form8}) completes the proof of Lemma 3 (\ref{Lemma3}). $\blacksquare$

{\bf \emph{Definition 3.}}
Let the trace-norm be defined as
\begin{equation}
T(\rho,\sigma):=\frac{1}{2}|\rho-\sigma|_1 = \frac{1}{2}\tr\{\sqrt{(\rho-\sigma)^2}\},
\end{equation}
and the Petz-Rényi relative entropy
\begin{equation}
D_\alpha (\rho||\sigma):=\frac{1}{\alpha -1} \ln \tr\{\rho^{\alpha} \sigma^{1-\alpha}\},
\end{equation}
for any $\alpha \in (0,1) \cup (1,\infty)$, with the symmetric version defined as
\begin{equation}
    \tilde{D}_\alpha := \frac{1}{2}(D_\alpha (\rho||\sigma) + D_\alpha (\sigma||\rho)).
\end{equation}

{\bf \emph{Theorem 1.}} {\it (Symmetric Petz-Rényi uncertainty relation) Let $\rho$ and $\sigma$ be any density matrices and $\hat{\theta}$ be any Hermitian operator. Then,}
\begin{equation}
\label{mainAppendix}
\tilde{D}_\alpha(\rho,\sigma) \geq \frac{1}{\alpha-1} \ln \frac{\cosh[(2\alpha-1)\arctanh(s(\rho,\sigma;\hat{\theta}))]}{\cosh[\arctanh(s(\rho,\sigma;\hat{\theta}))]},
\end{equation}
{\it where}
\begin{equation}
s(\rho,\sigma;\hat{\theta}):=\Big[\frac{(1/2)(\langle \hat{\theta} \rangle_\rho - \langle \hat{\theta} \rangle_\sigma)^2}{\langle \langle \hat{\theta} \rangle \rangle_\rho + \langle \langle \hat{\theta} \rangle \rangle_\sigma + (1/2)(\langle \hat{\theta} \rangle_\rho - \langle \hat{\theta} \rangle_\sigma)^2}\Big]^{1/2}.
\end{equation}
{\it Proof.} Using Definition 2, we have the following identity
\begin{equation}
\sum_{ij}P_{ij}^\alpha Q_{ij}^{1-\alpha} = \sum_{ij}p_i^\alpha q_j^{1-\alpha}|\langle p_i|q_j \rangle|^2 = \tr(\rho^\alpha \sigma^{1-\alpha}),
\end{equation}
which results in $D_\alpha(P|Q) = D_\alpha(\rho||\sigma)$, thus
\begin{equation}
\label{formB2appendix}
\tilde{D}_\alpha(P,Q) = \tilde{D}_\alpha(\rho,\sigma).
\end{equation}
We also have from Lemma 2 (\ref{Lemma2}) and Lemma 3 (\ref{Lemma3}),
\begin{equation}
\label{formB3}
\sqrt{\delta(P,Q)} \geq s(\rho,\sigma;\hat{\theta}).
\end{equation}
Finally, note that the function 
\begin{equation}
\label{formB4}
B(\alpha,x):=\frac{1}{\alpha-1} \ln \frac{\cosh[(2\alpha-1)\arctanh(x)]}{\cosh[\arctanh(x)]}
\end{equation}
is increasing for all $\alpha \in (0,1) \cup (1,\infty)$, which combined with (\ref{formB2appendix}), (\ref{formB3}) and Lemma 1 (\ref{Lemma1}), it results in
\begin{equation}
\tilde{D}_\alpha(\rho,\sigma)=\tilde{D}_\alpha(P,Q) \geq B(\alpha,\sqrt{\delta(P,Q)})\geq B(\alpha,s(\rho,\sigma;\hat{\theta})),
\end{equation}
which proves Theorem 1 (\ref{mainAppendix}). $\blacksquare$

{\bf \emph{Corollary 1.}} ({\it Generalized Holevo's inequality}) Let $\rho$ and $\sigma$ be any density matrices and $\hat{\theta}$ be any Hermitian operator. Then,
\begin{eqnarray}
\label{GenHolevosAppendix}
\tilde{D}_\alpha(\rho,\sigma) \geq \frac{1}{\alpha-1} \ln \frac{\cosh[(2\alpha-1)\arctanh(T(\rho,\sigma))]}{\cosh[\arctanh(T(\rho,\sigma))]}.
\end{eqnarray}

{\it Proof}. We use Theorem 1, we use a specific operator $\hat{\theta}=\hat{\omega}$ suggested in \cite{Salazar2024a} as follows, consider the following spectral decomposition of the operator $\rho-\sigma = \sum_k w_k |w_k \rangle \langle w_k |$. Now we define $\hat{\omega}=\hat{\omega}$ as
\begin{equation}
\label{form2}
\hat{\omega}:= \sum_{k, w_k \neq 0} \sign(w_k) |w_k\rangle \langle w_k|,
\end{equation}
where $\sign(x)=1~(-1)$, for $x>0~(x<0)$. Then, we write the trace-norm $|\rho-\sigma|_1$ in terms of $\hat{\omega}$ from (\ref{form2}),
\begin{equation}
\label{form3}
\langle \hat{\omega} \rangle_\rho - \langle \hat{\omega} \rangle_\sigma = \tr[\hat{\omega}(\rho-\sigma)]=\sum_k |w_k| =|\rho - \sigma|_1.
\end{equation}
Then, we observe that 
\begin{equation}
\label{form4}
\hat{\omega}^2 = \sum_{k,w_k \neq 0} \sign(w_k)^2 |w_k\rangle \langle w_k| = I - \hat{\epsilon},
\end{equation}
where $I$ is the identity operator and $\hat{\epsilon}:=\sum_{k,w_k=0}|w_k\rangle \langle w_k|$ , with averages
\begin{equation}
\label{form5}
\langle \hat{\epsilon} \rangle_\rho = \langle \hat{\epsilon} \rangle_\sigma := \epsilon,
\end{equation}
obtained from $\langle \hat{\epsilon}\rangle_\rho - \langle \hat{\epsilon}\rangle_\sigma = \tr[\hat{\epsilon}(\rho-\sigma)] = \sum_{k,w_k=0} w_k = 0$. We also have $0 \leq \epsilon \leq 1$, because $\rho, \sigma$ are positive definite and $\tr(\rho)=\tr(\sigma)=1$. From (\ref{form4}) and (\ref{form5}), we get
\begin{equation}
\label{form7}
\langle \hat{\omega}^2 \rangle_\rho = \langle \hat{\omega}^2\rangle_\sigma = 1 - \epsilon.
\end{equation}
Using the averages (\ref{form3}) and (\ref{form7}) we obtain
\begin{equation}
\label{form8}
s(\rho,\sigma;\hat{\omega})^2=\frac{(1/2)|\rho-\sigma|_1^2}{(2-2\epsilon)-\langle \hat{\omega}\rangle_\rho - \langle \hat{\omega}\rangle_\sigma + (1/2)|\rho-\sigma|_1^2}.
\end{equation}
Also check that
\begin{equation}
\label{form9}
 (1/2)|\rho - \sigma |_1^2 \leq \langle \hat{\omega} \rangle_\rho^2 + \langle \hat{\omega}\rangle_\sigma^2,
\end{equation}
directly from (\ref{form3}) and the expression $(1/2)(x-y)^2 \leq  (1/2)(x-y)^2 + (1/2)(x+y)^2 = x^2 + y^2$, for $x=\langle \hat{\omega} \rangle_\rho$, $y=\langle \hat{\omega} \rangle_\sigma$. In this case, we obtain from (\ref{form9}), $s(\rho,\sigma;\hat{\omega})^2 \geq \frac{|\rho-\sigma|_1^2}{4(1-\epsilon)} \geq \frac{|\rho-\sigma|_1^2}{4}$, which results in
\begin{equation}
\label{form10}
 s(\rho,\sigma;\hat{\omega}) \geq \frac{|\rho-\sigma|_1}{2}=T(\rho,\sigma).
\end{equation}
Now, using Theorem 1 with operator $\hat{\omega}$, we have $\tilde{D}_\alpha (\rho,\sigma) \geq B(\alpha, s(\rho,\sigma;\hat{\omega})) \geq B(\alpha, T(\rho,\alpha))
$, using (\ref{form10}) and the fact that $B(\alpha,x)$ is increasing, which proves the first inequality in (\ref{GenHolevosAppendix}). $\blacksquare$

{\bf \emph{Lemma 4.}} {\it The bound in Corollary 1 improves Pinsker's inequality.} 

{\it Proof}.
Let $T \in [0,1]$, then we show that
\begin{equation}
\label{Lemma4}
B(\alpha,T):=\frac{1}{\alpha-1} \ln \frac{\cosh[(2\alpha-1)\arctanh(T)]}{\cosh[\arctanh(T)]}
\geq
2\min(\alpha,1)T^2.
\end{equation}
For that purpose, we consider the function $g(\alpha,x)= \ln [\cosh((2\alpha-1)x)/\cosh(x)]$ for $x\geq 0$ and $\alpha \in (0,1) \cup (1, \infty)$. For a constant $x$, we have $\partial^2 g(\alpha,x)/\partial^2 \alpha = 4x^2\sech^2 [(x(1-2\alpha)] \geq 0$. Therefore, we could use the property of convex functions, $g(\alpha,x) - g(1,x) \geq g'(1,x)(\alpha-1)$, where $g'(\alpha,x)=\partial g(\alpha,x)/\partial \alpha$. Replacing $g'(1,x)=2 x \tanh(x)$ and $g(1,x)=0$, we have $g(\alpha,x) \geq 2 x \tanh(x) (\alpha-1)$, which yields for $\alpha>1$,
\begin{equation}
\label{appendixcorollary0}
B(\alpha,T) = \frac{1}{\alpha-1}g(\alpha,\arctanh(T)) \geq 2 T \arctanh(T) \geq 2T^2.
\end{equation}
Now for $0< \alpha <1$, we use a different strategy. First, note that $g(\alpha,x)$ can be conveniently rewritten as $g(\alpha,x)=2\arctanh(\tanh(\alpha x)\tanh((\alpha-1)x))$ using $\arctanh(x)=(1/2)\ln[(1+x)/(1-x)]$. Thus we obtain using $-\arctanh(x) \leq -x$,
\begin{equation}
\label{appendixcorollary1}
g(\alpha,x)  \leq -2\tanh(\alpha x)\tanh((1-\alpha)x) \leq 2\alpha(\alpha-1) \tanh(x)^2,
\end{equation}
where we used $\tanh(\epsilon x) \geq \epsilon \tanh(x)$ for $0< \epsilon < 1$ and $x>0$ (one can check this showing the function $j(x)=\tanh(\epsilon x)-\epsilon \tanh(x)$ is increasing in $x$ for any constant $\epsilon \in (0,1)$ and $j(0)=0$). Finally, we get from (\ref{appendixcorollary1}), using $\alpha -1 < 0$,
\begin{equation}
\label{appendixcorollary2}
B(\alpha,T)=\frac{1}{\alpha-1}g(\alpha,\arctanh(T)) \geq 2\alpha T^2.
\end{equation}
Combining (\ref{appendixcorollary0}) and (\ref{appendixcorollary2}), it results in $B(\alpha,T)\geq 2\min(\alpha,1)T^2$ for all $\alpha \geq 0$, which proves (\ref{Lemma4}).
$\blacksquare$

\bibliography{lib9}

\begin{thebibliography}{50}%
\makeatletter
\providecommand \@ifxundefined [1]{%
 \@ifx{#1\undefined}
}%
\providecommand \@ifnum [1]{%
 \ifnum #1\expandafter \@firstoftwo
 \else \expandafter \@secondoftwo
 \fi
}%
\providecommand \@ifx [1]{%
 \ifx #1\expandafter \@firstoftwo
 \else \expandafter \@secondoftwo
 \fi
}%
\providecommand \natexlab [1]{#1}%
\providecommand \enquote  [1]{``#1''}%
\providecommand \bibnamefont  [1]{#1}%
\providecommand \bibfnamefont [1]{#1}%
\providecommand \citenamefont [1]{#1}%
\providecommand \href@noop [0]{\@secondoftwo}%
\providecommand \href [0]{\begingroup \@sanitize@url \@href}%
\providecommand \@href[1]{\@@startlink{#1}\@@href}%
\providecommand \@@href[1]{\endgroup#1\@@endlink}%
\providecommand \@sanitize@url [0]{\catcode `\\12\catcode `\$12\catcode `\&12\catcode `\#12\catcode `\^12\catcode `\_12\catcode `\%12\relax}%
\providecommand \@@startlink[1]{}%
\providecommand \@@endlink[0]{}%
\providecommand \url  [0]{\begingroup\@sanitize@url \@url }%
\providecommand \@url [1]{\endgroup\@href {#1}{\urlprefix }}%
\providecommand \urlprefix  [0]{URL }%
\providecommand \Eprint [0]{\href }%
\providecommand \doibase [0]{http://dx.doi.org/}%
\providecommand \selectlanguage [0]{\@gobble}%
\providecommand \bibinfo  [0]{\@secondoftwo}%
\providecommand \bibfield  [0]{\@secondoftwo}%
\providecommand \translation [1]{[#1]}%
\providecommand \BibitemOpen [0]{}%
\providecommand \bibitemStop [0]{}%
\providecommand \bibitemNoStop [0]{.\EOS\space}%
\providecommand \EOS [0]{\spacefactor3000\relax}%
\providecommand \BibitemShut  [1]{\csname bibitem#1\endcsname}%
\let\auto@bib@innerbib\@empty
\bibitem [{\citenamefont {Holevo}(1972)}]{Holevo1972}%
  \BibitemOpen
  \bibfield  {author} {\bibinfo {author} {\bibfnamefont {A.~S.}\ \bibnamefont {Holevo}},\ }\href@noop {} {\bibfield  {journal} {\bibinfo  {journal} {Journal of Theoretical and Mathematical Physics}\ }\textbf {\bibinfo {volume} {13}},\ \bibinfo {pages} {1071} (\bibinfo {year} {1972})}\BibitemShut {NoStop}%
\bibitem [{\citenamefont {Wilde}(2018)}]{Wilde2018}%
  \BibitemOpen
  \bibfield  {author} {\bibinfo {author} {\bibfnamefont {M.}~\bibnamefont {Wilde}},\ }\href@noop {} {\bibfield  {journal} {\bibinfo  {journal} {IEEE International Symposium on Information Theory (ISIT)}\ ,\ \bibinfo {pages} {2331}} (\bibinfo {year} {2018})}\BibitemShut {NoStop}%
\bibitem [{\citenamefont {Iten}\ \emph {et~al.}(2017)\citenamefont {Iten}, \citenamefont {Renes},\ and\ \citenamefont {Sutter}}]{Iten2017}%
  \BibitemOpen
  \bibfield  {author} {\bibinfo {author} {\bibfnamefont {R.}~\bibnamefont {Iten}}, \bibinfo {author} {\bibfnamefont {J.~M.}\ \bibnamefont {Renes}}, \ and\ \bibinfo {author} {\bibfnamefont {D.}~\bibnamefont {Sutter}},\ }\href@noop {} {\bibfield  {journal} {\bibinfo  {journal} {IEEE Transactions on Information Theory}\ }\textbf {\bibinfo {volume} {63}},\ \bibinfo {pages} {1270} (\bibinfo {year} {2017})}\BibitemShut {NoStop}%
\bibitem [{\citenamefont {Ma}\ \emph {et~al.}(2008)\citenamefont {Ma}, \citenamefont {Zhang},\ and\ \citenamefont {Chen}}]{Zhihao2008}%
  \BibitemOpen
  \bibfield  {author} {\bibinfo {author} {\bibfnamefont {Z.}~\bibnamefont {Ma}}, \bibinfo {author} {\bibfnamefont {F.-L.}\ \bibnamefont {Zhang}}, \ and\ \bibinfo {author} {\bibfnamefont {J.-L.}\ \bibnamefont {Chen}},\ }\href {\doibase 10.1103/PhysRevA.78.064305} {\bibfield  {journal} {\bibinfo  {journal} {Phys. Rev. A}\ }\textbf {\bibinfo {volume} {78}},\ \bibinfo {pages} {64305} (\bibinfo {year} {2008})}\BibitemShut {NoStop}%
\bibitem [{\citenamefont {Albrecht}(1994)}]{Albrecht1994}%
  \BibitemOpen
  \bibfield  {author} {\bibinfo {author} {\bibfnamefont {A.}~\bibnamefont {Albrecht}},\ }\href {\doibase 10.1103/PhysRevD.50.2744} {\bibfield  {journal} {\bibinfo  {journal} {Phys. Rev. D}\ }\textbf {\bibinfo {volume} {50}},\ \bibinfo {pages} {2744} (\bibinfo {year} {1994})}\BibitemShut {NoStop}%
\bibitem [{\citenamefont {Luo}\ and\ \citenamefont {Zhang}(2004)}]{Shunlong2004}%
  \BibitemOpen
  \bibfield  {author} {\bibinfo {author} {\bibfnamefont {S.}~\bibnamefont {Luo}}\ and\ \bibinfo {author} {\bibfnamefont {Q.}~\bibnamefont {Zhang}},\ }\href {\doibase 10.1103/PhysRevA.69.032106} {\bibfield  {journal} {\bibinfo  {journal} {Phys. Rev. A}\ }\textbf {\bibinfo {volume} {69}},\ \bibinfo {pages} {32106} (\bibinfo {year} {2004})}\BibitemShut {NoStop}%
\bibitem [{\citenamefont {Kim}(2013)}]{Sejong2013}%
  \BibitemOpen
  \bibfield  {author} {\bibinfo {author} {\bibfnamefont {S.}~\bibnamefont {Kim}},\ }\href@noop {} {\bibfield  {journal} {\bibinfo  {journal} {Linear Algebra and its Applications}\ }\textbf {\bibinfo {volume} {438}},\ \bibinfo {pages} {2475} (\bibinfo {year} {2013})}\BibitemShut {NoStop}%
\bibitem [{\citenamefont {Audenaert}(2014)}]{Audenaert2014}%
  \BibitemOpen
  \bibfield  {author} {\bibinfo {author} {\bibfnamefont {K.~M.~R.}\ \bibnamefont {Audenaert}},\ }\href@noop {} {\bibfield  {journal} {\bibinfo  {journal} {Quantum Info. Comput.}\ }\textbf {\bibinfo {volume} {14}},\ \bibinfo {pages} {31} (\bibinfo {year} {2014})}\BibitemShut {NoStop}%
\bibitem [{\citenamefont {Petz}(1986)}]{PETZ198657}%
  \BibitemOpen
  \bibfield  {author} {\bibinfo {author} {\bibfnamefont {D.}~\bibnamefont {Petz}},\ }\href {\doibase https://doi.org/10.1016/0034-4877(86)90067-4} {\bibfield  {journal} {\bibinfo  {journal} {Reports on Mathematical Physics}\ }\textbf {\bibinfo {volume} {23}},\ \bibinfo {pages} {57} (\bibinfo {year} {1986})}\BibitemShut {NoStop}%
\bibitem [{\citenamefont {Buscemi}\ and\ \citenamefont {Datta}(2010)}]{Datta2010}%
  \BibitemOpen
  \bibfield  {author} {\bibinfo {author} {\bibfnamefont {F.}~\bibnamefont {Buscemi}}\ and\ \bibinfo {author} {\bibfnamefont {N.}~\bibnamefont {Datta}},\ }\href {\doibase 10.1109/TIT.2009.2039166} {\bibfield  {journal} {\bibinfo  {journal} {IEEE Transactions on Information Theory}\ }\textbf {\bibinfo {volume} {56}},\ \bibinfo {pages} {1447} (\bibinfo {year} {2010})}\BibitemShut {NoStop}%
\bibitem [{\citenamefont {Brandao}\ and\ \citenamefont {Datta}(2011)}]{Brandao2011IEEE}%
  \BibitemOpen
  \bibfield  {author} {\bibinfo {author} {\bibfnamefont {F.~G. S.~L.}\ \bibnamefont {Brandao}}\ and\ \bibinfo {author} {\bibfnamefont {N.}~\bibnamefont {Datta}},\ }\href {\doibase 10.1109/TIT.2011.2104531} {\bibfield  {journal} {\bibinfo  {journal} {IEEE Transactions on Information Theory}\ }\textbf {\bibinfo {volume} {57}},\ \bibinfo {pages} {1754} (\bibinfo {year} {2011})}\BibitemShut {NoStop}%
\bibitem [{\citenamefont {Kudler-Flam}(2023)}]{Kudler2023}%
  \BibitemOpen
  \bibfield  {author} {\bibinfo {author} {\bibfnamefont {J.}~\bibnamefont {Kudler-Flam}},\ }\href {\doibase 10.1103/PhysRevLett.130.021603} {\bibfield  {journal} {\bibinfo  {journal} {Phys. Rev. Lett.}\ }\textbf {\bibinfo {volume} {130}},\ \bibinfo {pages} {21603} (\bibinfo {year} {2023})}\BibitemShut {NoStop}%
\bibitem [{\citenamefont {Dechant}(2022)}]{Dechant2022}%
  \BibitemOpen
  \bibfield  {author} {\bibinfo {author} {\bibfnamefont {A.}~\bibnamefont {Dechant}},\ }\href {\doibase 10.1088/1751-8121/ac4ac0} {\bibfield  {journal} {\bibinfo  {journal} {Journal of Physics A: Mathematical and Theoretical}\ }\textbf {\bibinfo {volume} {55}},\ \bibinfo {pages} {94001} (\bibinfo {year} {2022})}\BibitemShut {NoStop}%
\bibitem [{\citenamefont {Vo}\ \emph {et~al.}(2022)\citenamefont {Vo}, \citenamefont {Vu},\ and\ \citenamefont {Hasegawa}}]{Vo_2022}%
  \BibitemOpen
  \bibfield  {author} {\bibinfo {author} {\bibfnamefont {V.~T.}\ \bibnamefont {Vo}}, \bibinfo {author} {\bibfnamefont {T.~V.}\ \bibnamefont {Vu}}, \ and\ \bibinfo {author} {\bibfnamefont {Y.}~\bibnamefont {Hasegawa}},\ }\href {\doibase 10.1088/1751-8121/ac9099} {\bibfield  {journal} {\bibinfo  {journal} {Journal of Physics A: Mathematical and Theoretical}\ }\textbf {\bibinfo {volume} {55}},\ \bibinfo {pages} {405004} (\bibinfo {year} {2022})}\BibitemShut {NoStop}%
\bibitem [{\citenamefont {Salazar}(2022{\natexlab{a}})}]{Salazar2022b}%
  \BibitemOpen
  \bibfield  {author} {\bibinfo {author} {\bibfnamefont {D.~S.~P.}\ \bibnamefont {Salazar}},\ }\href {\doibase 10.1103/PhysRevE.106.L032101} {\bibfield  {journal} {\bibinfo  {journal} {Phys. Rev. E}\ }\textbf {\bibinfo {volume} {106}},\ \bibinfo {pages} {L032101} (\bibinfo {year} {2022}{\natexlab{a}})}\BibitemShut {NoStop}%
\bibitem [{\citenamefont {Barato}\ and\ \citenamefont {Seifert}(2015)}]{Barato2015A}%
  \BibitemOpen
  \bibfield  {author} {\bibinfo {author} {\bibfnamefont {A.~C.}\ \bibnamefont {Barato}}\ and\ \bibinfo {author} {\bibfnamefont {U.}~\bibnamefont {Seifert}},\ }\href {\doibase 10.1103/PhysRevLett.114.158101} {\bibfield  {journal} {\bibinfo  {journal} {Physical Review Letters}\ }\textbf {\bibinfo {volume} {114}},\ \bibinfo {pages} {158101} (\bibinfo {year} {2015})}\BibitemShut {NoStop}%
\bibitem [{\citenamefont {Gingrich}\ \emph {et~al.}(2016)\citenamefont {Gingrich}, \citenamefont {Horowitz}, \citenamefont {Perunov},\ and\ \citenamefont {England}}]{Gingrich2016}%
  \BibitemOpen
  \bibfield  {author} {\bibinfo {author} {\bibfnamefont {T.~R.}\ \bibnamefont {Gingrich}}, \bibinfo {author} {\bibfnamefont {J.~M.}\ \bibnamefont {Horowitz}}, \bibinfo {author} {\bibfnamefont {N.}~\bibnamefont {Perunov}}, \ and\ \bibinfo {author} {\bibfnamefont {J.~L.}\ \bibnamefont {England}},\ }\href {\doibase 10.1103/PhysRevLett.116.120601} {\bibfield  {journal} {\bibinfo  {journal} {Physical Review Letters}\ }\textbf {\bibinfo {volume} {116}},\ \bibinfo {pages} {120601} (\bibinfo {year} {2016})},\ \Eprint {http://arxiv.org/abs/1512.02212} {arXiv:1512.02212} \BibitemShut {NoStop}%
\bibitem [{\citenamefont {Polettini}\ \emph {et~al.}(2016)\citenamefont {Polettini}, \citenamefont {Lazarescu},\ and\ \citenamefont {Esposito}}]{Polettini2017}%
  \BibitemOpen
  \bibfield  {author} {\bibinfo {author} {\bibfnamefont {M.}~\bibnamefont {Polettini}}, \bibinfo {author} {\bibfnamefont {A.}~\bibnamefont {Lazarescu}}, \ and\ \bibinfo {author} {\bibfnamefont {M.}~\bibnamefont {Esposito}},\ }\href {\doibase 10.1103/PhysRevE.94.052104} {\bibfield  {journal} {\bibinfo  {journal} {Phys. Rev. E}\ }\textbf {\bibinfo {volume} {94}},\ \bibinfo {pages} {052104} (\bibinfo {year} {2016})}\BibitemShut {NoStop}%
\bibitem [{\citenamefont {Pietzonka}\ and\ \citenamefont {Seifert}(2017)}]{Pietzonka2017}%
  \BibitemOpen
  \bibfield  {author} {\bibinfo {author} {\bibfnamefont {P.}~\bibnamefont {Pietzonka}}\ and\ \bibinfo {author} {\bibfnamefont {U.}~\bibnamefont {Seifert}},\ }\href {\doibase 10.1103/PhysRevLett.120.190602} {\bibfield  {journal} {\bibinfo  {journal} {Physical Review Letters}\ }\textbf {\bibinfo {volume} {120}},\ \bibinfo {pages} {190602} (\bibinfo {year} {2017})},\ \Eprint {http://arxiv.org/abs/1705.05817} {arXiv:1705.05817} \BibitemShut {NoStop}%
\bibitem [{\citenamefont {Hasegawa}\ and\ \citenamefont {{Van Vu}}(2019{\natexlab{a}})}]{Hasegawa2019}%
  \BibitemOpen
  \bibfield  {author} {\bibinfo {author} {\bibfnamefont {Y.}~\bibnamefont {Hasegawa}}\ and\ \bibinfo {author} {\bibfnamefont {T.}~\bibnamefont {{Van Vu}}},\ }\href {\doibase 10.1103/PhysRevLett.123.110602} {\bibfield  {journal} {\bibinfo  {journal} {Phys. Rev. Lett.}\ }\textbf {\bibinfo {volume} {123}},\ \bibinfo {pages} {110602} (\bibinfo {year} {2019}{\natexlab{a}})}\BibitemShut {NoStop}%
\bibitem [{\citenamefont {Hasegawa}\ and\ \citenamefont {{Van Vu}}(2019{\natexlab{b}})}]{Hasegawa2019b}%
  \BibitemOpen
  \bibfield  {author} {\bibinfo {author} {\bibfnamefont {Y.}~\bibnamefont {Hasegawa}}\ and\ \bibinfo {author} {\bibfnamefont {T.}~\bibnamefont {{Van Vu}}},\ }\href@noop {} {\bibfield  {journal} {\bibinfo  {journal} {Phys. Rev. E}\ }\textbf {\bibinfo {volume} {99}} (\bibinfo {year} {2019}{\natexlab{b}})}\BibitemShut {NoStop}%
\bibitem [{\citenamefont {Vo}\ \emph {et~al.}(2020)\citenamefont {Vo}, \citenamefont {{Van Vu}},\ and\ \citenamefont {Hasegawa}}]{VanTuan2020}%
  \BibitemOpen
  \bibfield  {author} {\bibinfo {author} {\bibfnamefont {V.~T.}\ \bibnamefont {Vo}}, \bibinfo {author} {\bibfnamefont {T.}~\bibnamefont {{Van Vu}}}, \ and\ \bibinfo {author} {\bibfnamefont {Y.}~\bibnamefont {Hasegawa}},\ }\href {\doibase 10.1103/PhysRevE.102.062132} {\bibfield  {journal} {\bibinfo  {journal} {Phys. Rev. E}\ }\textbf {\bibinfo {volume} {102}},\ \bibinfo {pages} {62132} (\bibinfo {year} {2020})}\BibitemShut {NoStop}%
\bibitem [{\citenamefont {Vu}\ and\ \citenamefont {Hasegawa}(2020)}]{Van_Vu_2020}%
  \BibitemOpen
  \bibfield  {author} {\bibinfo {author} {\bibfnamefont {T.~V.}\ \bibnamefont {Vu}}\ and\ \bibinfo {author} {\bibfnamefont {Y.}~\bibnamefont {Hasegawa}},\ }\href {\doibase 10.1088/1751-8121/ab64a4} {\bibfield  {journal} {\bibinfo  {journal} {Journal of Physics A: Mathematical and Theoretical}\ }\textbf {\bibinfo {volume} {53}},\ \bibinfo {pages} {75001} (\bibinfo {year} {2020})}\BibitemShut {NoStop}%
\bibitem [{\citenamefont {Timpanaro}\ \emph {et~al.}(2019)\citenamefont {Timpanaro}, \citenamefont {Guarnieri}, \citenamefont {Goold},\ and\ \citenamefont {Landi}}]{Timpanaro2019B}%
  \BibitemOpen
  \bibfield  {author} {\bibinfo {author} {\bibfnamefont {A.~M.}\ \bibnamefont {Timpanaro}}, \bibinfo {author} {\bibfnamefont {G.}~\bibnamefont {Guarnieri}}, \bibinfo {author} {\bibfnamefont {J.}~\bibnamefont {Goold}}, \ and\ \bibinfo {author} {\bibfnamefont {G.~T.}\ \bibnamefont {Landi}},\ }\href {\doibase 10.1103/PhysRevLett.123.090604} {\bibfield  {journal} {\bibinfo  {journal} {Physical Review Letters}\ }\textbf {\bibinfo {volume} {123}},\ \bibinfo {pages} {090604} (\bibinfo {year} {2019})},\ \Eprint {http://arxiv.org/abs/1904.07574} {arXiv:1904.07574} \BibitemShut {NoStop}%
\bibitem [{\citenamefont {Liu}\ \emph {et~al.}(2020)\citenamefont {Liu}, \citenamefont {Gong},\ and\ \citenamefont {Ueda}}]{Liu2020}%
  \BibitemOpen
  \bibfield  {author} {\bibinfo {author} {\bibfnamefont {K.}~\bibnamefont {Liu}}, \bibinfo {author} {\bibfnamefont {Z.}~\bibnamefont {Gong}}, \ and\ \bibinfo {author} {\bibfnamefont {M.}~\bibnamefont {Ueda}},\ }\href {\doibase 10.1103/PhysRevLett.125.140602} {\bibfield  {journal} {\bibinfo  {journal} {Phys. Rev. Lett.}\ }\textbf {\bibinfo {volume} {125}},\ \bibinfo {pages} {140602} (\bibinfo {year} {2020})}\BibitemShut {NoStop}%
\bibitem [{\citenamefont {Horowitz}\ and\ \citenamefont {Gingrich}(2020)}]{Horowitz2022}%
  \BibitemOpen
  \bibfield  {author} {\bibinfo {author} {\bibfnamefont {J.~M.}\ \bibnamefont {Horowitz}}\ and\ \bibinfo {author} {\bibfnamefont {T.~R.}\ \bibnamefont {Gingrich}},\ }\href {\doibase 10.1038/s41567-019-0702-6} {\bibfield  {journal} {\bibinfo  {journal} {Nature Physics}\ }\textbf {\bibinfo {volume} {16}},\ \bibinfo {pages} {15} (\bibinfo {year} {2020})}\BibitemShut {NoStop}%
\bibitem [{\citenamefont {Potts}\ and\ \citenamefont {Samuelsson}(2019)}]{Potts2019}%
  \BibitemOpen
  \bibfield  {author} {\bibinfo {author} {\bibfnamefont {P.~P.}\ \bibnamefont {Potts}}\ and\ \bibinfo {author} {\bibfnamefont {P.}~\bibnamefont {Samuelsson}},\ }\href {http://arxiv.org/abs/1904.04913} {\bibfield  {journal} {\bibinfo  {journal} {Phys. Rev. E}\ }\textbf {\bibinfo {volume} {100}},\ \bibinfo {pages} {052137} (\bibinfo {year} {2019})},\ \Eprint {http://arxiv.org/abs/1904.04913} {arXiv:1904.04913} \BibitemShut {NoStop}%
\bibitem [{\citenamefont {Proesmans}\ and\ \citenamefont {Horowitz}(2019)}]{Proesmans2019}%
  \BibitemOpen
  \bibfield  {author} {\bibinfo {author} {\bibfnamefont {K.}~\bibnamefont {Proesmans}}\ and\ \bibinfo {author} {\bibfnamefont {J.}~\bibnamefont {Horowitz}},\ }\href {http://arxiv.org/abs/1902.07008} {\bibfield  {journal} {\bibinfo  {journal} {Journal of Statistical Mechanics: Theory and Experiment}\ }\textbf {\bibinfo {volume} {5}},\ \bibinfo {pages} {054005} (\bibinfo {year} {2019})},\ \Eprint {http://arxiv.org/abs/1902.07008} {arXiv:1902.07008} \BibitemShut {NoStop}%
\bibitem [{\citenamefont {Francica}(2022)}]{Gianluca2022}%
  \BibitemOpen
  \bibfield  {author} {\bibinfo {author} {\bibfnamefont {G.}~\bibnamefont {Francica}},\ }\href {\doibase 10.1103/PhysRevE.105.014129} {\bibfield  {journal} {\bibinfo  {journal} {Phys. Rev. E}\ }\textbf {\bibinfo {volume} {105}},\ \bibinfo {pages} {14129} (\bibinfo {year} {2022})}\BibitemShut {NoStop}%
\bibitem [{\citenamefont {Salazar}(2022{\natexlab{b}})}]{Salazar2022d}%
  \BibitemOpen
  \bibfield  {author} {\bibinfo {author} {\bibfnamefont {D.~S.~P.}\ \bibnamefont {Salazar}},\ }\href {\doibase 10.1103/PhysRevE.106.L062104} {\bibfield  {journal} {\bibinfo  {journal} {Phys. Rev. E}\ }\textbf {\bibinfo {volume} {106}},\ \bibinfo {pages} {L062104} (\bibinfo {year} {2022}{\natexlab{b}})}\BibitemShut {NoStop}%
\bibitem [{\citenamefont {Brandner}\ \emph {et~al.}(2018)\citenamefont {Brandner}, \citenamefont {Hanazato},\ and\ \citenamefont {Saito}}]{Brandner2018}%
  \BibitemOpen
  \bibfield  {author} {\bibinfo {author} {\bibfnamefont {K.}~\bibnamefont {Brandner}}, \bibinfo {author} {\bibfnamefont {T.}~\bibnamefont {Hanazato}}, \ and\ \bibinfo {author} {\bibfnamefont {K.}~\bibnamefont {Saito}},\ }\href@noop {} {\bibfield  {journal} {\bibinfo  {journal} {Phys. Rev. Lett.}\ }\textbf {\bibinfo {volume} {120}} (\bibinfo {year} {2018})}\BibitemShut {NoStop}%
\bibitem [{\citenamefont {Carollo}\ \emph {et~al.}(2019)\citenamefont {Carollo}, \citenamefont {Jack},\ and\ \citenamefont {Garrahan}}]{Carollo2019}%
  \BibitemOpen
  \bibfield  {author} {\bibinfo {author} {\bibfnamefont {F.}~\bibnamefont {Carollo}}, \bibinfo {author} {\bibfnamefont {R.~L.}\ \bibnamefont {Jack}}, \ and\ \bibinfo {author} {\bibfnamefont {J.~P.}\ \bibnamefont {Garrahan}},\ }\href {\doibase 10.1103/PhysRevLett.122.130605} {\bibfield  {journal} {\bibinfo  {journal} {Physical Review Letters}\ }\textbf {\bibinfo {volume} {122}},\ \bibinfo {pages} {130605} (\bibinfo {year} {2019})},\ \Eprint {http://arxiv.org/abs/arXiv:1811.04969v1} {arXiv:arXiv:1811.04969v1} \BibitemShut {NoStop}%
\bibitem [{\citenamefont {Liu}\ and\ \citenamefont {Segal}(2019)}]{Liu2019}%
  \BibitemOpen
  \bibfield  {author} {\bibinfo {author} {\bibfnamefont {J.}~\bibnamefont {Liu}}\ and\ \bibinfo {author} {\bibfnamefont {D.}~\bibnamefont {Segal}},\ }\href {http://arxiv.org/abs/1904.11963} {\bibfield  {journal} {\bibinfo  {journal} {Phys. Rev. E}\ }\textbf {\bibinfo {volume} {99}},\ \bibinfo {pages} {1} (\bibinfo {year} {2019})},\ \Eprint {http://arxiv.org/abs/1904.11963} {arXiv:1904.11963} \BibitemShut {NoStop}%
\bibitem [{\citenamefont {{Van Vu}}\ and\ \citenamefont {Saito}(2022)}]{VanVu2022b}%
  \BibitemOpen
  \bibfield  {author} {\bibinfo {author} {\bibfnamefont {T.}~\bibnamefont {{Van Vu}}}\ and\ \bibinfo {author} {\bibfnamefont {K.}~\bibnamefont {Saito}},\ }\href {\doibase 10.1103/PhysRevLett.128.140602} {\bibfield  {journal} {\bibinfo  {journal} {Phys. Rev. Lett.}\ }\textbf {\bibinfo {volume} {128}},\ \bibinfo {pages} {140602} (\bibinfo {year} {2022})}\BibitemShut {NoStop}%
\bibitem [{\citenamefont {Miller}\ \emph {et~al.}(2021)\citenamefont {Miller}, \citenamefont {Mohammady}, \citenamefont {Perarnau-Llobet},\ and\ \citenamefont {Guarnieri}}]{Miller2021}%
  \BibitemOpen
  \bibfield  {author} {\bibinfo {author} {\bibfnamefont {H.~J.~D.}\ \bibnamefont {Miller}}, \bibinfo {author} {\bibfnamefont {M.~H.}\ \bibnamefont {Mohammady}}, \bibinfo {author} {\bibfnamefont {M.}~\bibnamefont {Perarnau-Llobet}}, \ and\ \bibinfo {author} {\bibfnamefont {G.}~\bibnamefont {Guarnieri}},\ }\href {\doibase 10.1103/PhysRevLett.126.210603} {\bibfield  {journal} {\bibinfo  {journal} {Phys. Rev. Lett.}\ }\textbf {\bibinfo {volume} {126}},\ \bibinfo {pages} {210603} (\bibinfo {year} {2021})}\BibitemShut {NoStop}%
\bibitem [{\citenamefont {Pires}\ \emph {et~al.}(2021)\citenamefont {Pires}, \citenamefont {Modi},\ and\ \citenamefont {C{\'{e}}leri}}]{Pires2021}%
  \BibitemOpen
  \bibfield  {author} {\bibinfo {author} {\bibfnamefont {D.~P.}\ \bibnamefont {Pires}}, \bibinfo {author} {\bibfnamefont {K.}~\bibnamefont {Modi}}, \ and\ \bibinfo {author} {\bibfnamefont {L.~C.}\ \bibnamefont {C{\'{e}}leri}},\ }\href {\doibase 10.1103/PhysRevE.103.032105} {\bibfield  {journal} {\bibinfo  {journal} {Phys. Rev. E}\ }\textbf {\bibinfo {volume} {103}},\ \bibinfo {pages} {32105} (\bibinfo {year} {2021})}\BibitemShut {NoStop}%
\bibitem [{\citenamefont {Guarnieri}\ \emph {et~al.}(2019)\citenamefont {Guarnieri}, \citenamefont {Landi}, \citenamefont {Clark},\ and\ \citenamefont {Goold}}]{Guarnieri2019}%
  \BibitemOpen
  \bibfield  {author} {\bibinfo {author} {\bibfnamefont {G.}~\bibnamefont {Guarnieri}}, \bibinfo {author} {\bibfnamefont {G.~T.}\ \bibnamefont {Landi}}, \bibinfo {author} {\bibfnamefont {S.~R.}\ \bibnamefont {Clark}}, \ and\ \bibinfo {author} {\bibfnamefont {J.}~\bibnamefont {Goold}},\ }\href {https://journals.aps.org/prresearch/pdf/10.1103/PhysRevResearch.1.033021} {\bibfield  {journal} {\bibinfo  {journal} {Physical Review Research}\ }\textbf {\bibinfo {volume} {1}},\ \bibinfo {pages} {33021} (\bibinfo {year} {2019})},\ \Eprint {http://arxiv.org/abs/1901.10428v2} {arXiv:1901.10428v2} \BibitemShut {NoStop}%
\bibitem [{\citenamefont {Hasegawa}(2020)}]{Hasegawa2019a}%
  \BibitemOpen
  \bibfield  {author} {\bibinfo {author} {\bibfnamefont {Y.}~\bibnamefont {Hasegawa}},\ }\href {\doibase 10.1103/PhysRevLett.125.050601} {\bibfield  {journal} {\bibinfo  {journal} {Phys. Rev. Lett.}\ }\textbf {\bibinfo {volume} {125}},\ \bibinfo {pages} {50601} (\bibinfo {year} {2020})}\BibitemShut {NoStop}%
\bibitem [{\citenamefont {Hasegawa}(2021{\natexlab{a}})}]{hasegawa2021}%
  \BibitemOpen
  \bibfield  {author} {\bibinfo {author} {\bibfnamefont {Y.}~\bibnamefont {Hasegawa}},\ }\href {\doibase 10.1103/PhysRevLett.127.240602} {\bibfield  {journal} {\bibinfo  {journal} {Phys. Rev. Lett.}\ }\textbf {\bibinfo {volume} {127}},\ \bibinfo {pages} {240602} (\bibinfo {year} {2021}{\natexlab{a}})}\BibitemShut {NoStop}%
\bibitem [{\citenamefont {Hasegawa}(2023)}]{Hasegawa2023}%
  \BibitemOpen
  \bibfield  {author} {\bibinfo {author} {\bibfnamefont {Y.}~\bibnamefont {Hasegawa}},\ }\href {\doibase 10.1038/s41467-023-38074-8} {\bibfield  {journal} {\bibinfo  {journal} {Nature Communications}\ }\textbf {\bibinfo {volume} {14}},\ \bibinfo {pages} {2828} (\bibinfo {year} {2023})}\BibitemShut {NoStop}%
\bibitem [{\citenamefont {Hasegawa}(2021{\natexlab{b}})}]{Hasegawa2021b}%
  \BibitemOpen
  \bibfield  {author} {\bibinfo {author} {\bibfnamefont {Y.}~\bibnamefont {Hasegawa}},\ }\href {\doibase 10.1103/PhysRevLett.126.010602} {\bibfield  {journal} {\bibinfo  {journal} {Phys. Rev. Lett.}\ }\textbf {\bibinfo {volume} {126}},\ \bibinfo {pages} {10602} (\bibinfo {year} {2021}{\natexlab{b}})}\BibitemShut {NoStop}%
\bibitem [{\citenamefont {Nussbaum}\ and\ \citenamefont {Szko{\l}a}(2009)}]{Nussbaum2009}%
  \BibitemOpen
  \bibfield  {author} {\bibinfo {author} {\bibfnamefont {M.}~\bibnamefont {Nussbaum}}\ and\ \bibinfo {author} {\bibfnamefont {A.}~\bibnamefont {Szko{\l}a}},\ }\href {http://www.jstor.org/stable/30243657} {\bibfield  {journal} {\bibinfo  {journal} {The Annals of Statistics}\ }\textbf {\bibinfo {volume} {37}},\ \bibinfo {pages} {1040} (\bibinfo {year} {2009})}\BibitemShut {NoStop}%
\bibitem [{\citenamefont {Datta}\ \emph {et~al.}(2013)\citenamefont {Datta}, \citenamefont {Mosonyi}, \citenamefont {Hsieh},\ and\ \citenamefont {Brand{\~{a}}o}}]{Datta2013}%
  \BibitemOpen
  \bibfield  {author} {\bibinfo {author} {\bibfnamefont {N.}~\bibnamefont {Datta}}, \bibinfo {author} {\bibfnamefont {M.}~\bibnamefont {Mosonyi}}, \bibinfo {author} {\bibfnamefont {M.-H.}\ \bibnamefont {Hsieh}}, \ and\ \bibinfo {author} {\bibfnamefont {F.~G. S.~L.}\ \bibnamefont {Brand{\~{a}}o}},\ }\href {\doibase 10.1109/TIT.2013.2282160} {\bibfield  {journal} {\bibinfo  {journal} {IEEE Transactions on Information Theory}\ }\textbf {\bibinfo {volume} {59}},\ \bibinfo {pages} {8014} (\bibinfo {year} {2013})}\BibitemShut {NoStop}%
\bibitem [{\citenamefont {Salazar}(2024)}]{Salazar2023d}%
  \BibitemOpen
  \bibfield  {author} {\bibinfo {author} {\bibfnamefont {D.~S.~P.}\ \bibnamefont {Salazar}},\ }\href {\doibase 10.1103/PhysRevE.109.L012103} {\bibfield  {journal} {\bibinfo  {journal} {Phys. Rev. E}\ }\textbf {\bibinfo {volume} {109}},\ \bibinfo {pages} {L012103} (\bibinfo {year} {2024})}\BibitemShut {NoStop}%
\bibitem [{\citenamefont {Androulakis}\ and\ \citenamefont {John}(2023{\natexlab{a}})}]{Androulakis2023}%
  \BibitemOpen
  \bibfield  {author} {\bibinfo {author} {\bibfnamefont {G.}~\bibnamefont {Androulakis}}\ and\ \bibinfo {author} {\bibfnamefont {T.~C.}\ \bibnamefont {John}},\ }\href {\doibase 10.1142/S0219025723500212} {\bibfield  {journal} {\bibinfo  {journal} {Infinite Dimensional Analysis, Quantum Probability and Related Topics}\ }\textbf {\bibinfo {volume} {0}},\ \bibinfo {pages} {2350021} (\bibinfo {year} {2023}{\natexlab{a}})}\BibitemShut {NoStop}%
\bibitem [{\citenamefont {Androulakis}\ and\ \citenamefont {John}(2023{\natexlab{b}})}]{Androulakis2023b}%
  \BibitemOpen
  \bibfield  {author} {\bibinfo {author} {\bibfnamefont {G.}~\bibnamefont {Androulakis}}\ and\ \bibinfo {author} {\bibfnamefont {T.~C.}\ \bibnamefont {John}},\ }\href {\doibase 10.1142/S0129055X23600024} {\bibfield  {journal} {\bibinfo  {journal} {Reviews in Mathematical Physics}\ }\textbf {\bibinfo {volume} {0}},\ \bibinfo {pages} {2360002} (\bibinfo {year} {2023}{\natexlab{b}})}\BibitemShut {NoStop}%
\bibitem [{\citenamefont {Nishiyama}(2022)}]{Nishiyama2022b}%
  \BibitemOpen
  \bibfield  {author} {\bibinfo {author} {\bibfnamefont {T.}~\bibnamefont {Nishiyama}},\ }\href@noop {} {\bibfield  {journal} {\bibinfo  {journal} {https://arxiv.org/abs/2210.09571}\ } (\bibinfo {year} {2022})}\BibitemShut {NoStop}%
\bibitem [{\citenamefont {Salazar}(2023{\natexlab{a}})}]{Salazar2024a}%
  \BibitemOpen
  \bibfield  {author} {\bibinfo {author} {\bibfnamefont {D.}~\bibnamefont {Salazar}},\ }\href@noop {} {\bibfield  {journal} {\bibinfo  {journal} {https://arxiv.org/abs/2311.13536}\ } (\bibinfo {year} {2023}{\natexlab{a}})}\BibitemShut {NoStop}%
\bibitem [{\citenamefont {Salazar}(2023{\natexlab{b}})}]{Salazar2023a}%
  \BibitemOpen
  \bibfield  {author} {\bibinfo {author} {\bibfnamefont {D.~S.~P.}\ \bibnamefont {Salazar}},\ }\href {https://journals.aps.org/pre/accepted/8f079R75O6cE711091822080f42846a87d813a6c6} {\bibfield  {journal} {\bibinfo  {journal} {Physical Review E}\ }\textbf {\bibinfo {volume} {107}},\ \bibinfo {pages} {L062103} (\bibinfo {year} {2023}{\natexlab{b}})}\BibitemShut {NoStop}%
\bibitem [{\citenamefont {Falasco}\ \emph {et~al.}(2022)\citenamefont {Falasco}, \citenamefont {Esposito},\ and\ \citenamefont {Delvenne}}]{Falasco_2022}%
  \BibitemOpen
  \bibfield  {author} {\bibinfo {author} {\bibfnamefont {G.}~\bibnamefont {Falasco}}, \bibinfo {author} {\bibfnamefont {M.}~\bibnamefont {Esposito}}, \ and\ \bibinfo {author} {\bibfnamefont {J.-C.}\ \bibnamefont {Delvenne}},\ }\href {\doibase 10.1088/1751-8121/ac52e2} {\bibfield  {journal} {\bibinfo  {journal} {Journal of Physics A: Mathematical and Theoretical}\ }\textbf {\bibinfo {volume} {55}},\ \bibinfo {pages} {124002} (\bibinfo {year} {2022})}\BibitemShut {NoStop}%
\end{thebibliography}%

\end{document}